\documentclass[aps,10pt,twocolumn,showpacs,pra,citeautoscript,amsmath,amssymb,floatfix,nofootinbib,superscriptaddress,longbibliography]{revtex4-2}
\usepackage[left=2cm,right=2cm,top=2cm,bottom=2cm]{geometry}

\usepackage[utf8]{inputenc}
\usepackage[english]{babel}
\usepackage[T1]{fontenc}
\usepackage{xcolor}
\usepackage{hyperref}
\usepackage{cleveref} %ref (1-3)
\usepackage{mathtools}
\usepackage{verbatim}%block comment
\usepackage{tikz-cd} %diagrams
\usepackage{amsmath}
\usepackage{amsfonts}
\usepackage{amssymb}
\usepackage{cancel}
\usepackage{float}      %utilizar [H] em imagens
\usepackage{braket}
\usepackage{array}   % for \newcolumntype macro
\newcolumntype{L}{>{$}l<{$}} % math-mode version of "l" column type

\usepackage{bbold}

\usepackage{amsthm}

\usepackage{enumerate}
\usepackage{braket}

\usepackage{framed}
\usepackage{graphicx}

\usepackage{natbib}

\usepackage{csquotes}

\usepackage{xcolor}
\usepackage[normalem]{ulem}
\usepackage{soul}

\begin{document}
\title{Can multiple observers detect KS-contextuality?}
\author{Arthur C. R. Dutra*}
\affiliation{Instituto de Matemática, Estatística e Computação Científica,
Universidade Estadual de Campinas, Campinas, SP, Brazil}
\author{Roberto D. Baldij\~ao}
\affiliation{Institute of Physics, University of São Paulo, R. do Matão 1371, São Paulo 05508-090, SP, Brazil}
\affiliation{International Centre for Theory of Quantum Technologies, University of Gda\'{n}sk, 80-309 Gda\'{n}sk, Poland}

\author{Marcelo Terra Cunha}
\affiliation{Instituto de Matemática, Estatística e Computação Científica,
Universidade Estadual de Campinas, Campinas, SP, Brazil}

\begin{abstract}
KS-contextuality is a crucial feature of quantum theory. Previous research demonstrated the vanishing of $N$-cycle KS-contextuality in setups where multiple independent observers measure sequentially on the same system, which we call Public Systems. This phenomenon can be explained as the additional observers' measurements degrading the state and depleting the quantum resource. This explanation would imply that state-independent contextuality should survive in such a system. In this paper, we show that this is not the case. We achieved this result by simulating an observer trying to violate the Peres-Mermin noncontextuality inequality in a Public System. Additionally, we provide an analytical description of our setup, explaining the loss of contextuality even in the state-independent case. Ultimately, these results show that state-independent contextuality is not independent of what happens to the system in-between the measurements of a context.
\end{abstract}

\maketitle
\onecolumngrid

\twocolumngrid
{
  \renewcommand{\thefootnote}%
    {\fnsymbol{footnote}}
  \footnotetext[1]{arthur.couto.oliveira@alumni.usp.br}
}

\section{Introduction}

Quantum physics has many counter-intuitive features that deeply contrast with our everyday experience. One of such features, essential for understanding nonclassicality in quantum theory, is KS-contextuality,
which is also a vital resource in many quantum information tasks \cite{howard2014contextuality,PhysRevA.72.012325,PhysRevA.55.4089}.

Classical reasoning presupposes that all measurements on a given state possess a preexisting value. It attributes any apparent randomness to an incomplete description of the state, which one could reinterpret as a statistical average of more precisely defined states \cite{RevModPhys.38.447}. Many physicists conjectured that it might be possible to complete quantum theory with additional variables, an idea famously associated with Einstein, Podolsky, and Rosen \cite{einstein1935can}. 
In this complete description of quantum theory, any measurement would have a predetermined outcome. 

KS-contextuality concerns the possibility of assigning these predetermined values independently of context, i.e., independent of which other (compatible) measurements are performed jointly.
This feature is arguably a signature of classicality and hence expected of models that want to give a classical account of quantum theory. 
When such an assignment is possible, we say the
a theory is KS-\emph{non}contextual; if it is \emph{in}possible, then it is KS-contextual. 

Kochen and Specker showed that quantum theory predicts correlations that cannot be reproduced by a KS-noncontextual theory, a result known as the  Kochen-Specker theorem \cite{kochen1990problem}. The classical correlations, produced by noncontextual predetermined outcomes, can be translated into noncontextuality inequalities, establishing bounds that any KS-noncontextual statistics must respect. Thus, we can detect KS-noncontextuality via violations of such inequalities.

 These inequalities also allow us to distinguish two types of contextuality: if any state produces violations (including the maximally mixed state), we call it state-\emph{independent} contextuality (SIC), but if only some states do so, we say it is state-\emph{dependent} contextuality. A comprehensive review of KS-contextuality can be found in Ref.~\cite{budroni2021quantumReview}. 

Since KS-contextuality is a nonclassical phenomenon, exploring processes that cause the emergence of KS-\emph{non}contextuality is an interesting endeavor. Such classical limits contribute to a deeper comprehension of the foundations of quantum theory and might help to prevent classicality from manifesting itself in applications that rely on this quantum resource.

Ref.~\cite{Baldijao_2020_StateDependentMultiple} showed that in a Public System (PS)-- systems shared by multiple independent observers that measure sequentially -- observers lose the capability to detect state-dependent KS-contextuality as the number of observers increases. Ref.~\cite{Baldijao_2020_StateDependentMultiple} attributes this classical limit to state-degradation: since these inequalities are pretty sensitive to the state, it could no longer violate the inequality after being altered by other observers' measurements.

 This explanation, which focuses on the state as harboring the quantum resource, leads naturally to the question: Would state-\emph{in}dependent contextuality survive in a Public System? This seems straightforwardly true: since any state violates SIC inequalities, degrading the state should not affect KS-contextuality.

In this paper, we numerically simulate an observer trying to violate the Peres-Mermin state-independent inequality \cite{PhysRevLett.101.210401} in a PS. We show that the presence of multiple observers can still vanquish state-independent contextuality--proving that, in this case, state degradation cannot be the driving factor behind its destruction.

Our analytical description then explains the disappearance of SIC violations: even though SIC is independent of the system's state before the measurements of a context, SIC is not oblivious to what happens in-between the measurements. We can understand this in two ways: In the Schrödinger picture, the effect of the additional observers is equivalent to a depolarization channel that effectively break the correlations among measurements of a context. That is, re-preparing the state between the measurements kills the possibility of witnessing contextuality. In the Heisenberg picture, where the state does not change due to the dynamics imposed by the additional observers, the measurements are altered, also affecting the ability to witness SIC-inequalities violations. 

Our results also shed light on why actual experiments of the Peres-Mermin square never reach the value expected by quantum theory for \emph{any} state \cite{PhysRevLett.103.160405}: noise introduced to the state in-between the measurements can also diminish SIC violations. Our work may also be of interest for future applications in quantum information tasks, where a state is subject to public access.

The paper is organized as follows: in  Section~\ref{PSS}, we explain public systems; Section~\ref{Background} explains the Peres-Mermin scenario; Section~\ref{protocol} describes the implementation of the public system in our simulation; Section~\ref{Results} exhibits the simulation and analytical results, and in Section~\ref{Discussion} we discuss our findings and present our results. Appendix~\ref{Appendix} expands on calculating the expected values of sequential measurements.

\section{Public Systems (PS)}\label{PSS}

The usual way to test a noncontextuality inequality is
with a single observer performing various measurement rounds to get the necessary statistics. Each round, they prepare a quantum state, measure a series of compatible observables, register the outcomes, discard
the system, and repeat. An illustrative example of this protocol is the experiment described in Ref.~\cite{PhysRevLett.103.160405} with photons.

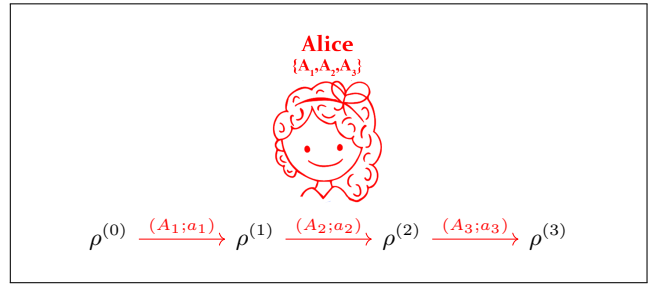
\begin{figure}[]
\begin{framed}
\centering
   \includegraphics[width=.3\linewidth]{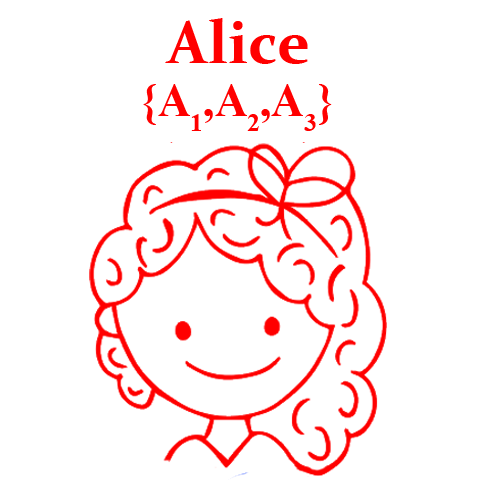}

\begin{minipage}{\textwidth}  %%<--- here
\centering
        \begin{tikzcd}[column sep={6em,between origins}]
            \rho^{(0)}\ar[red]{r}{\color{red}(A_{1};a_{1})}  &   \rho^{(1)}\ar[red]{r}{\color{red}(A_{2};a_{2})}   & \rho^{(2)}\ar[red]{r}{\color{red}(A_{3};a_{3})} & \rho^{(3)}
        \end{tikzcd}
\end{minipage}
\end{framed}
    \caption[Sequential measurement round with a single observer.]{A diagrammatic representation of a sequential measurement round with a single observer. Time flows from left to right, and each arrow represents a measurement - it starts on the state before the measurement and points to the output state. The pair \((A_{i}; a_i)\) above the arrows indicates the measured observable and the measurement result.}
    \label{fig: single observer}
\end{figure}

We will use this initial protocol to introduce our diagrammatic representation of a sequential round of measurements to detect KS-contextuality, shown in Fig.~\ref{fig: single observer}. We represent each measurement by an arrow that starts on the pre-measurement state \(\rho^{(i)}\) and points to the post-measurement state \(\rho^{(i+1)}\). We labeled the arrows with the ordered pair \((A_{k}; a_k)\), which indicates which measurement $A_k$ was performed and the result $a_k$ obtained. Measurements are color-coded to the observer who implemented them.

This single-observer setting assumes the observer entirely controls the system. In general, that is not the case. Indeed, one possible generalization is to consider more observers that probe the systems, particularly setups in which multiple independent observers have full access to the system, which we call Public Systems (PS).

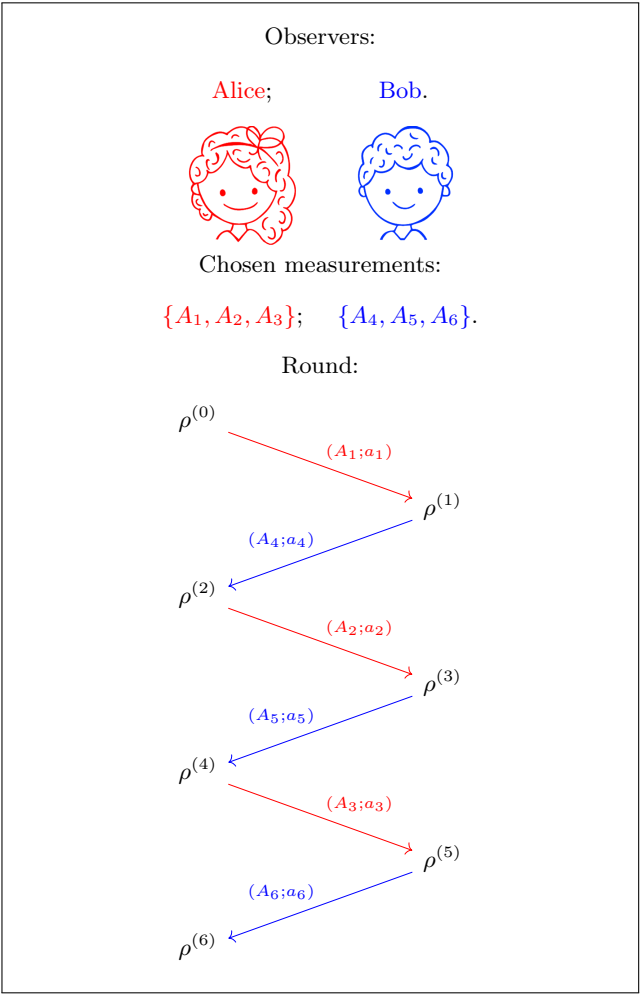
\begin{figure}
\centering
\begin{framed}
\begin{minipage}{0.9\linewidth}
\begin{center}
Observers:     
\end{center}
\textcolor{red}{Alice}; \hspace{1.2cm} \textcolor{blue}{Bob}.
\begin{center}
\includegraphics[width=.6\linewidth]{"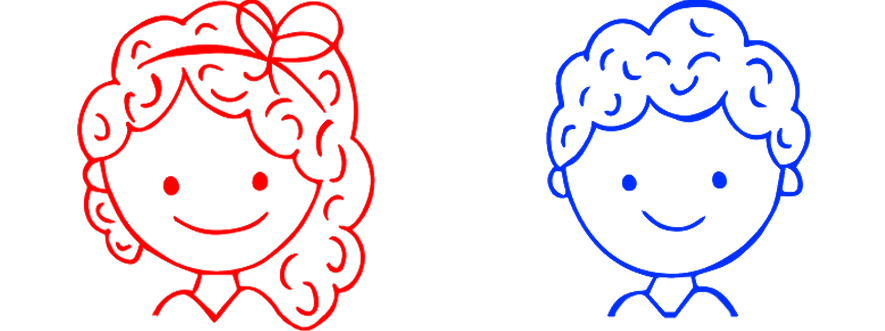"}

Chosen measurements:
\[\textcolor{red}{\{ A_{1}, A_{2}, A_{3}\}};\hspace{.4cm}\textcolor{blue}{\{ A_{4}, A_{5}, A_{6}\}}.\]

Round:

\end{center}

\begin{tikzcd}[column sep={10em,between origins}]
\rho^{(0)}\arrow[red]{rd} {\color{red}(A_{1};a_{1})} & \\
& \rho^{(1)} \arrow[blue]{ld}[swap]{\color{blue}(A_{4};a_4)}\\ 
\rho^{(2)}\arrow[red]{rd}{\color{red}(A_{2};a_{2})} &\\
& \rho^{(3)} \arrow[blue]{ld}[swap]{\color{blue}(A_{5};a_{5})}\\ \rho^{(4)}\arrow[red]{rd} {\color{red}(A_{3
};a_{3})} & \\
& \rho^{(5)} \arrow[blue]{ld}[swap]{\color{blue}(A_{6};a_{6})} \\
\rho^{(6)}&
\end{tikzcd}

\end{minipage}
\end{framed}
\caption[Public System measurement round.]{A diagrammatic representation of a PS measurement round with two observers. Time flows from top to bottom, and each arrow represents a measurement - it starts on the state before the measurement and points to the output state - and is color-coded to the observer who implemented it. The pair \((A_{i}; a_i)\) above the arrows indicates the measured observable and the measurement result.}
\label{fig: multiple observer ex}
\end{figure}

In Fig.~\ref{fig: multiple observer ex}, we depict a round in a PS with observers Alice and Bob. Alice wants to measure the context \(\{A_1,A_2,A_3\}\) and Bob wishes to implement \(\{A_4,A_5,A_6\}\). Each time Alice accesses the system, she finds it at state \(\rho^{(i)}\), on the left side of the image, and gets to implement only one of her measurements. On the flip side, Bob gets each post-measurement state from Alice, \(\rho^{(i+1)}\) on the right side of the image and makes one of his measurements.

Of course, the example in Fig.~\ref{fig: multiple observer ex} can be generalized to $N$ additional observers simply by adding additional columns on the diagram for each new observer. In a PS round, numerous independent observers may take turns accessing the system and performing measurements after the initial state preparation. As they operate independently, it is crucial to note that the measurements conducted by one observer may not necessarily be compatible with those of the others.

Even without this nomenclature, PSs have been explored in the literature; for instance, previous research examined steering \cite{PhysRevA.98.012305}, Bell nonlocality \cite{PhysRevLett.114.250401,PhysRevA.99.022305}, preparation contextuality~\cite{Anwer2021noiserobust} and KS-contextuality \cite{Baldijao_2020_StateDependentMultiple}. 

Ref. \cite{Baldijao_2020_StateDependentMultiple} tested if numerous observers could violate odd $N$-cycle inequalities \cite{PhysRevA.88.022118}--a family of state-dependent inequalities, such as the KCBS inequality \cite{PhysRevLett.101.020403}--under various implementations of PS and found that only a limited number of observers could detect violations.
Their findings point to state degradation as the reason for the emergence of KS-noncontextuality. As the $N$-cycle inequalities are pretty sensitive to the state used for the violation, it is reasonable to attribute this phenomenon to the multiple observers degrading the state, thereby depleting the quantum resource. This reasoning would imply that SIC would survive in a PS since any state could produce a violation. 

This work falsifies this hypothesis using two qubits as our PS and the Peres-Mermin measurements which we describe below.

\section{The Peres-Mermin Square}
\label{Background}

When studying contextuality, we delimit scenarios where observers can only implement a few predetermined measurements. One of the simplest scenarios in which we can observe SIC is the Peres-Mermin (PM) square \cite{PERES1990107,PhysRevLett.65.3373,APeres_1991}. In this scenario, the observer has access to nine dichotomic measurements that can be arranged in a square grid:
\begin{align}
\begin{split}
    &A_{11} \hspace{0.3cm} A_{12} \hspace{0.3cm} A_{13} \\
    &A_{21} \hspace{0.3cm} A_{22} \hspace{0.3cm} A_{23} \\
    &A_{31} \hspace{0.3cm} A_{32} \hspace{0.3cm} A_{33}. \\
\end{split}
\end{align}
Each of them can output either $-1$ or $1$. These represent nine different attributes that can either be present or absent. 
The contexts of this scenario are composed of the rows and columns of the square:
\begin{subequations}
\label{contexts}
\begin{equation}
\label{c+i}
R_1\coloneqq\{A_{11},A_{12},A_{13}\},
\end{equation}
\begin{equation}
R_2\coloneqq\{A_{21},A_{22},A_{23}\},
\end{equation}
\begin{equation}
\label{c+f}
R_3\coloneqq\{A_{31},A_{32},A_{33}\},
\end{equation}
\begin{equation}
\label{c-i}
C_1\coloneqq\{A_{11},A_{21},A_{31}\},
\end{equation}
\begin{equation}
C_2\coloneqq\{A_{12},A_{22},A_{32}\},
\end{equation}
\begin{equation}
\label{c-f}
C_3\coloneqq\{A_{13},A_{23},A_{33}\}.
\end{equation}
\end{subequations}

If we assume KS-noncontextuality, i.e., assume that we  may attribute a deterministic value \(v(A_{ij})\in\{-1,1\}\) to each measurement independently of context, we are constrained by the following noncontextuality inequality~\cite{PhysRevLett.101.210401}:
\begin{align}
\begin{split}
    &\Sigma\coloneqq\langle A_{11}A_{12}A_{13} \rangle +\langle A_{21}A_{22}A_{23}\rangle +\langle A_{31}A_{32}A_{33} \rangle\\ 
&-\langle A_{11}A_{21}A_{31}\rangle-\langle A_{12}A_{22}A_{32}\rangle -\langle A_{13}A_{23}A_{33}\rangle\stackrel{NC}{\leq} 4,
\end{split}
\label{inequality}
\end{align}
where the superscript `NC' stands for noncontextual. 

In a two-qubit system, one can physically implement this scenario using the following observables
\footnote{There are $10$ implementations of the Peres-Mermin square using Pauli observables on two qubits \cite{holweck2021testing}\cite{saniga2007veldkamp}, we choose the most symmetric for our implementation.}:
\begin{align}
\label{squareMeasurements}
\begin{split}
A_{11}&=\sigma_y\otimes \sigma_z \hspace{0.3 cm} A_{12}=\sigma_z\otimes \sigma_y \hspace{0.3 cm}
A_{13}=\sigma_x\otimes \sigma_x \\
A_{21}&=\sigma_z\otimes \sigma_x \hspace{0.3 cm} A_{22}=\sigma_x\otimes \sigma_z \hspace{0.3 cm}
A_{23}=\sigma_y\otimes \sigma_y \\
A_{31}&=\sigma_x\otimes \sigma_y \hspace{0.3 cm} A_{32}=\sigma_y\otimes \sigma_x \hspace{0.3 cm}
A_{33}=\sigma_z\otimes \sigma_z,
\end{split}
\end{align}
where \(\sigma_x\), \(\sigma_y\), \(\sigma_z\) are the Pauli spin matrices.

Note that the product of the observables from each row, Eqs. \eqref{c+i} to \eqref{c+f}, is \(I\) while the product of the ones from the columns, Eqs. \eqref{c-i} to \eqref{c-f}, is \(-I\). This implies that measuring \(\Sigma\) will result in a violation of the noncontextual quota~\eqref{inequality} \emph{regardless of the state} \(\rho\) used in the test\footnote{The fact that the operators in each context multiply to the identity is behind the state-independence in any SIC inequality\cite{cabello2015necessary}.}:
\begin{align}
\begin{split}
\Sigma\stackrel{Q}{=}6\langle I \rangle_\rho=6,
\label{qequality}
\end{split}
\end{align}
where `Q' stands for `quantum', as this equality emerges from the quantum description. 

Experiments demonstrating violations of Ineq.~\eqref{inequality} through a physical implementation of the PM scenario have been successfully conducted using ions \cite{kirchmair2009state}, photons \cite{PhysRevLett.103.160405}, and nuclear magnetic resonance \cite{PhysRevLett.104.160501}; nonetheless, there is still room for progress in approaching the theoretical quantum value of $\Sigma=6$.

Equality~\eqref{qequality} might lead to the belief that, when dealing with SIC, what happens to the state is irrelevant, even in multiple observers setup. As we will see in the following sections, that is not the case.

\section{Protocol for the Peres-Mermin Square in a PS}
\label{protocol}

We simulated a PS of two qubits where one observer is trying to violate the PM noncontextuality inequality~\eqref{inequality} (the Main Observer), while $N$ other observers are simply measuring aimlessly (Passersby Observers)\footnote{It would also make sense to consider the situation in which all observers tried to violate the inequality. However, the action of the Passerby Observers is arguably more suited to PS, where the other observers may interact with the system for different purposes. We ultimately choose to simulate this setup since this approach provides the additional advantage of allowing for a straightforward analytical description, as will be demonstrated in the subsequent sections.}. In this scenario, it is natural to constrain the measurements to the PM scenario. 

In our simulation, during each round, all observers adhere to the prescribed protocol:
\begin{enumerate}
    \item The system is prepared in state \(\rho^{(0)}\);
    \item An order of access to the system is randomly decided among the $N+1$ observers;
    \item Observers randomly pick their measurements - one context~\eqref{contexts} for the Main Observer and three observables from~\eqref{squareMeasurements} for each Passerby Observer;
    \item The Main Observer randomly selects an order of implementation for the selected context;
    \item Observers take turns accessing the system. In each access, they perform a single measurement on the state coming from the last observer's measurement;
    \item Main Observer tallies the results obtained for their chosen context.
\end{enumerate}
After many rounds, the Main Observer calculates the average value for each context to obtain $\Sigma$ (see Ineq.~\eqref{inequality}).
As one can see, in each simulation we define an initial state \(\rho^{(0)}\) - to be prepared at the start of every round - and the number of Passerby Observers \((N)\).

We calculate the post-measurement state using the standard state-update rule for projective measurements on mixed states \cite[p. 99]{nielsen_chuang_2010}
\begin{align}
    \rho^{(i+1)} = \frac{\Pi^{o}_k \rho^{(i)} \Pi^{o}_k}{{\rm tr}[\Pi^{o}_k \rho^{(i)} \Pi^{o}_k]},
    \label{eq: stateUpdate}
\end{align}
that describes the post-measurement state $\rho^{(i+1)}$ when the initial state was $\rho^{(i)}$ and the outcome $o$, associated with the projection $\Pi^{o}_k$, was obtained from measuring $A_k$.

\begin{figure}
\begin{framed}
\hspace{.4cm} Main Observer: \hspace{.4cm}Passerby Observer:

\hspace{0.2cm}\textcolor{red}{Alice}; \hspace{1.5cm} \textcolor{blue}{Bob}.

\begin{center}
\hspace{.4cm}\includegraphics[width=.6\linewidth]{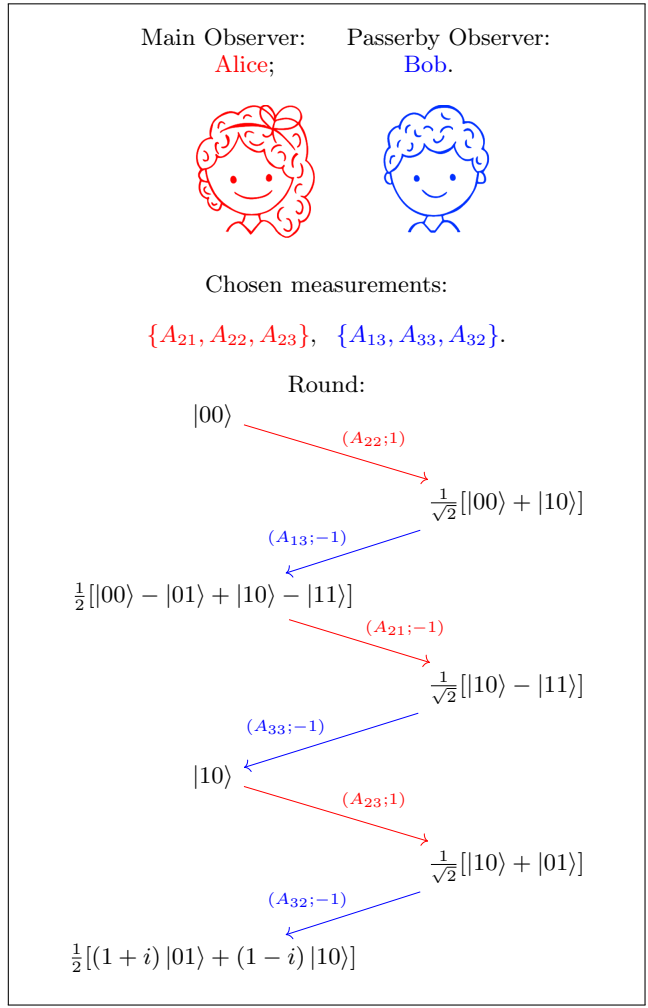}
  
\end{center}

Chosen measurements:
\[\textcolor{red}{\{ A_{21}, A_{22}, A_{23}\}},\hspace{.2cm}\textcolor{blue}{\{ A_{13}, A_{33}, A_{32}\}}.\]

Round:

\begin{tikzcd}[column sep={12em,between origins}]
\ket{00}\arrow[red]{rd} {\color{red}(A_{22};1)} & \\
& \frac{1}{\sqrt{2}}[\ket{00}+\ket{10}]\arrow[blue,swap]{ld}{\color{blue}(A_{13};-1)}\\ 
\frac{1}{2}[\ket{00}-\ket{01}+\ket{10}-\ket{11}]\arrow[red]{rd}{\color{red}(A_{21};-1)} &\\
& \frac{1}{\sqrt{2}}[\ket{10}-\ket{11}] \arrow[blue,swap]{ld}{\color{blue}(A_{33};-1)}\\ \ket{10} \arrow[red]{rd} {\color{red}(A_{23
};1)} & \\
& \frac{1}{\sqrt{2}}[\ket{10}+\ket{01}] \arrow[blue,swap]{ld}{\color{blue}(A_{32};-1)} \\
\frac{1}{2}[(1+i)\ket{01}+(1-i)\ket{10}]&
\end{tikzcd}

\end{framed}
\caption{A diagrammatic representation of a measurement round in the PS following our protocol. Main Observer Alice chooses a context to implement, while Passerby Observer Bob chooses three random observables without regard for compatibility. They then take turns implementing those measurements - represented in the diagram by the color-coded arrows. The pair \((A_{ij}; a_{ij})\) above the arrows indicates the measured observable and the measurement result.}
\label{fig: PM example}
\end{figure}

Fig.~\ref{fig: PM example} shows an example of a round in our PS with the Main Observer (Alice) and only one Passerby Observer~(Bob). The initial state is \(\rho^{(0)}=\ket{00}\bra{00}\). The kets \(\{\ket{0},\ket{1}\}\) are the eigenstates of the \(\sigma_z\) Pauli observable.

Notice that, in each round, the observers randomly pick an order of access, their measurements, and an order of implementation. In the round depicted in Fig.~\ref{fig: PM example} Alice is the first to access the system and measures context \(R_2=\{ A_{21}, A_{22}, A_{23}\}\), while Bob is the second to access it and chooses the measurements \(\{ A_{13}, A_{33}, A_{32}\}\) which, as is the case, need not be compatible. 

\section{Results} \label{Results}

\begin{figure}
    \includegraphics[width=\linewidth]{"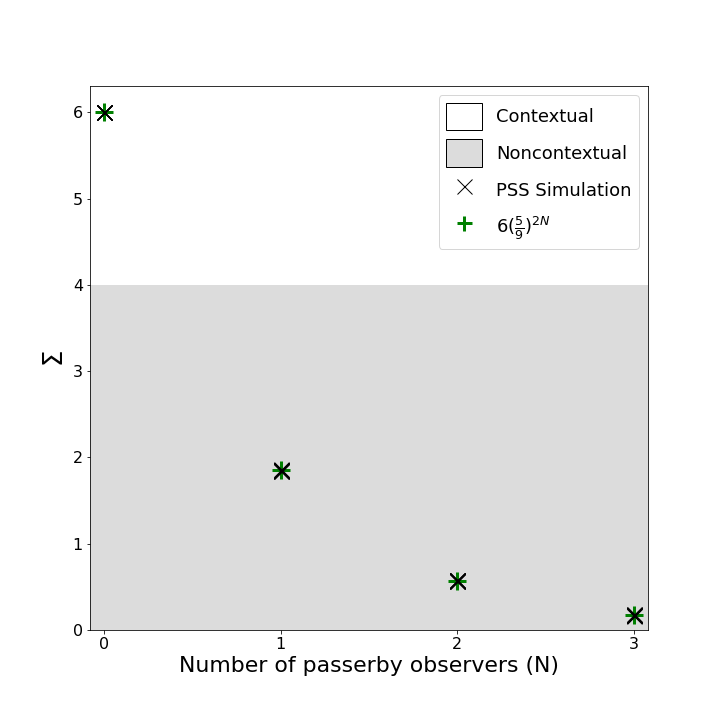"}
    \caption{\(\Sigma\) as a function of the number of Passerby Observers after a million rounds in our PS. The shaded region of the graph corresponds to values that are compatible with KS-noncontextuality.}
    \label{PPS-figure}
\end{figure}

We wrote a python code\footnote{The code is available on the GitHub repository \url{https://github.com/turdutra/PS_simulation/}.} to simulate the protocol described in Sec.\ref{protocol} to assess if SIC would persist in our PS. For each number of passerby observers \((N\in\{1,2,3\})\), we chose ten random pure state as the initial state \(\rho^{(0)}\) and ran the simulation spanning one million rounds. The results, represented by the values of \(\Sigma\) (See Ineq.~\eqref{inequality}), are in Fig.~\ref{PPS-figure}.

Our findings reveal that the PS can wash away KS-contextuality, even in the state-independent case. Moreover, in the PM scenario, just one Passerby Observer is enough to prevent violations of the noncontextual quota~\eqref{inequality}.

One interesting aspect of the protocol we implemented in our PS (see Sec.~\ref{protocol}) is that Passersby Observers behave stochastically; each of their measurements is independent of the last. This independence allows us to model their influence using a Markovian quantum channel, which provides an analytical framework for understanding what is happening to KS-contextuality.

From here on, we will take the perspective of the Main Observer and interpret our PS as an open quantum system. Let us begin considering a system with just one Passerby Observer.

We denote by $\Gamma_{ij}$ the map induced by the Passerby Observer implementing measurement $A_{ij}$.
Every observable from Eqs.~\eqref{squareMeasurements} can be decomposed as a difference of projectors into the subspaces of the $\pm1$ eigenvalues
\begin{align}
    A_{ij}=\Pi^+_{ij}-\Pi^-_{ij}.
\end{align}
Using the state-update rule~\eqref{eq: stateUpdate}, the state after the measurement--when its result is unknown--is
\begin{align}
    \begin{split}    \Gamma_{ij}(\rho)&=tr(\Pi_{ij}^+\rho)\frac{\Pi_{ij}^+\rho\Pi_{ij}^+}        {tr(\Pi_{ij}^+\rho)}+tr(\Pi_{ij}^-\rho)\frac{\Pi_{ij}^-\rho\Pi_{ij}^-}    {tr(\Pi_{ij}^-\rho)}=\\
        &=\Pi_{ij}^+\rho\Pi_{ij}^++\Pi_{ij}^-\rho\Pi_{ij}^-,
    \end{split}
\end{align}
which can also be written as \footnote{Note that $\Pi_{ij}^\pm=\frac{1}{2}(I\pm A_{ij})$.}
\begin{align}
\begin{split}
    \Gamma_{ij}(\rho)&=\Pi_{ij}^+\rho\Pi_{ij}^++\Pi_{ij}^-\rho\Pi_{ij}^-=\\
    &=\frac{1}{2}\rho + \frac{1}{2}A_{ij}\rho A_{ij}.
\end{split}
\label{pre-canal}
\end{align}

 Between each of the Main Observer's measurements, the state could have been subject to any of the nine measurements of the PM square~\eqref{squareMeasurements}. As a result of this interference, Main Observer receives a mixed state for their subsequent measurement, a convex combination of the action of each $\Gamma_{ij}$--we will denote the average measurement map by $\Gamma$. Since each $A_{ij}$ is chosen uniformly at random by the Passerby Observer, we get
\begin{align}
    \begin{split}
        \Gamma(\rho) = \frac{1}{2}\rho+\frac{1}{18}\sum_{ij}A_{ij}\rho A_{ij}.
       \label{canal}
    \end{split}
\end{align}

From the perspective of the Main observer, who is constrained to measuring PM observables, channel~\eqref{canal} is equivalent to the depolarization channel:
\begin{equation}
    \Gamma(\rho)=\frac{5}{9}\rho+\frac{1}{9} I,
    \label{canal depolarização}
\end{equation}
as shown in Appendix~\ref{appendix depolarzation}.

\begin{figure}
\begin{framed}
\begin{minipage}{0.9\linewidth}
\hspace{.4cm} Main Observer: \hspace{.4cm}Passerby Observer:

\hspace{0.2cm}\textcolor{red}{Alice}; \hspace{1.5cm} \textcolor{blue}{Bob}.

\begin{center}
\hspace{.4cm}\includegraphics[width=.6\linewidth]{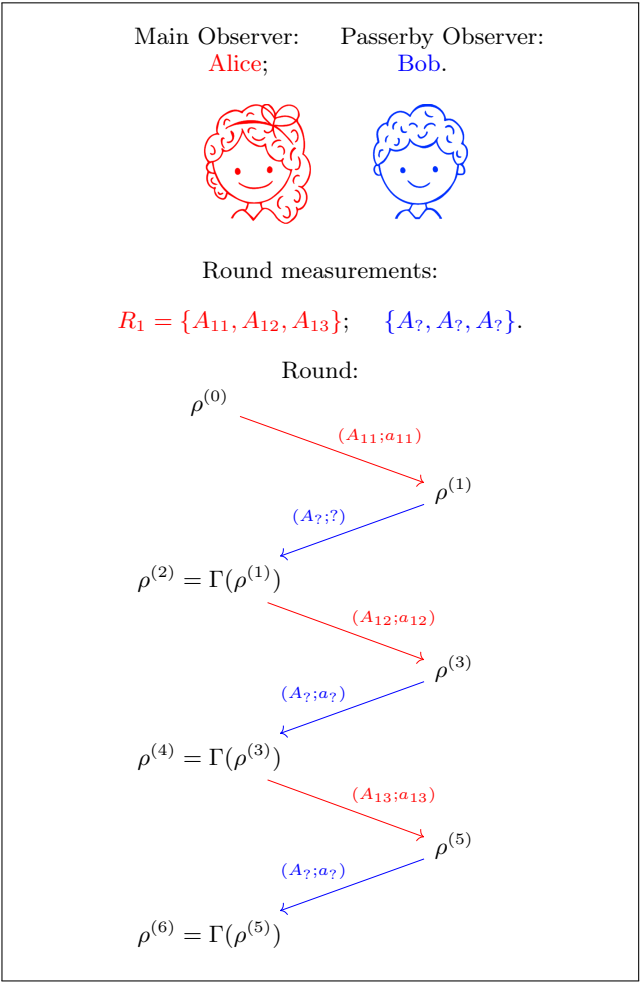}
  
\end{center}

Round measurements:
\[\textcolor{red}{R_1=\{ A_{11}, A_{12}, A_{13}\}};\hspace{.4cm}\textcolor{blue}{\{ A_{?}, A_{?}, A_{?}\}}.\]

Round:

\hspace*{-.5cm} % * makes sure space is not discarted
\begin{tikzcd}[column sep={10em,between origins}]
\rho^{(0)}\arrow[red]{rd} {\color{red}(A_{11};a_{11})} & \\
& \rho^{(1)} \arrow[blue]{ld}[swap]{\color{blue}(A_{?};?)}\\ 
\rho^{(2)}=\Gamma(\rho^{(1)})\arrow[red]{rd}{\color{red}(A_{12};a_{12})} &\\
& \rho^{(3)} \arrow[blue]{ld}[swap]{\color{blue}(A_{?};a_{?})}\\ \rho^{(4)}=\Gamma(\rho^{(3)})\arrow[red]{rd} {\color{red}(A_{13
};a_{13})} & \\
& \rho^{(5)} \arrow[blue]{ld}[swap]{\color{blue}(A_{?};a_{?})} \\
\rho^{(6)}=\Gamma(\rho^{(5)})&
\end{tikzcd}

\end{minipage}
\end{framed}
\caption{A diagrammatic representation of a PS round with two observers from the perspective of the Main Observer, Alice. After each of her measurements, she knows that Bob implemented one of the PM measurements but does not know which one or its result, represented by \((A_?,a_?)\). We can model Bob's influence with the quantum channel \(\Gamma\).}
\label{fig: open system}
\end{figure}

This dynamics is illustrated in Fig.~\ref{fig: open system}, with Main Observer Alice and Passerby Observer Bob, where the states on which Alice implements her second and third measurements are disturbed by channel $\Gamma$~\eqref{canal depolarização}.

%%%%%%%%%%%%%%%%%%%%%%%%%
%%% Previous version %%%
%%%%%%%%%%%%%%%%%%%%%%%%
%Following numerous rounds under the influence of the Passerby Observer, if the Main Observer calculates the average value of a context, such as $R_1$, the channel of Eq.~\eqref{canal depolarização} will impact {\color{red} its \sout{the average} value \sout{(as detailed in Appendix~\ref{Appendix})}}:
%
%\begin{align}
%\begin{split}
%& \langle A_{11}A_{12}A_{13} \rangle_{\rho^{(0)}} \stackrel{\Gamma}{\rightarrow} \\
%& \langle A_{11}[{\rho^{(0)}}] \cdot A_{12}[{\Gamma(\rho^{(1)})}] \cdot A_{13}[{\Gamma(\rho^{(3)})}] \rangle.
%\end{split}
\label{eq: muilti time measurment}
%\end{align}
%
%{\color{red} The last term in Eq.~\eqref{eq: muilti time measurment} is not an expected value of an observable, but rather it is a three-point correlation function. Operationally, it is the average, over many rounds, of the product of outcomes of the three measurements (each acting on different states). Since $\Gamma$ alters the post-measurement state, this object is in general different from the usual expected value (see Appendix~\ref{Appendix}). Indeed, one can see that}
%the identity portion of channel $\Gamma$~\eqref{canal depolarização} will hinder the correlations between the Main Observer's measurements.

Following numerous rounds under the influence of the Passerby Observer, if the Main Observer calculates the average value of a context, such as $R_1$, the channel of Eq.~\eqref{canal depolarização} will impact the results. 
%While 
Indeed, under no action of the Passerby, one can identify:
\begin{equation}
 \langle A_{11}A_{12}A_{13} \rangle_{\rho^{(0)}}
= \langle A_{11}[\rho^{(0)}] \cdot A_{12}[\rho^{(1)}] \cdot A_{13}[\rho^{(2)}] \rangle,
\end{equation}
where $\rho^{(1)}$ is the updated state coming from $\rho^{(0)}$ and the result of measurement $A_{11}$ (and   similarly for $\rho^{(2)}$) and the brackets just indicate a mean value estimation of such three-point correlation function. Now, if we consider
the Passerby presence and actions,  the multi-time aspect of these correlations becomes more explicit. 
Now,
\begin{align}
\begin{split}
& \langle A_{11}A_{12}A_{13} \rangle_{\rho^{(0)}}
\stackrel{\Gamma}{\rightarrow} \\
& \langle A_{11}[{\rho^{(0)}}] \cdot A_{12}[{\Gamma(\rho^{(1)})}] \cdot A_{13}[{\Gamma(\rho^{(3)})}] \rangle.
\end{split}
\label{eq: muilti time measurment}
\end{align}
This means that Alice's best effort to estimate the mean value of a context, under the same strategy she has always used, will now be a good estimator for a
 three-point correlation function. 
 Operationally, it is still the average, over many rounds, of the product of outcomes of the three compatible measurements, however now each acting on different states. 
 Since $\Gamma$ alters the post-measurement state, this object is in general different from the usual expected value (see Appendix~\ref{Appendix}). Indeed, one can see that
the depolarizing portion of channel $\Gamma$ in Eq.~\eqref{canal depolarização} will hinder the correlations between the Main Observer's measurements.

 Generalizing the above to a PS with $N$ Passerby Observers, their measurements  have an effect equivalent to the channel
\begin{equation}
    \Gamma^N(\rho)=\left(\frac{5}{9}\right)^N \rho+\frac{1-\left(\frac{5}{9}\right)^N}{4} I.
    \label{eq: multi observers channel}
\end{equation}

From Eq.~\eqref{eq: multi observers channel}, we see that, if we take the limit of \(N\rightarrow\infty\), channel $\Gamma^N$~\eqref{eq: multi observers channel} will completely decorrelate the outcomes of the Main Observer's measurements. Denoting by $\rho*=\frac{1}{4}I$ the maximally mixed state, we have:
\begin{align}
\begin{split}
    &\lim_{N\rightarrow \infty}\langle A_{11}[{\rho^{(0)}}] \cdot A_{12}[{\Gamma^N(\rho^{(1)})}] \cdot A_{13}[{\Gamma^N(\rho^{(3)})}] \rangle=\\
        &=\langle A_{11}[\rho^{(0)}] \cdot A_{12}[\rho*]\cdot A_{13}[\rho*]\rangle=\\
        &=\langle A_{11}\rangle_{\rho^{(0)}} \langle A_{12}\rangle_{\rho*}\langle A_{13}\rangle_{\rho*}=0.
\end{split}
\end{align}

Notice that, even if the initial state \(\rho^{(0)}\) is already the maximally mixed state, a fixed point of $\Gamma^N$, the correlations are still destroyed, since 
\begin{equation}
    1=\langle A_{11}A_{12}A_{13}\rangle_{\rho*}\neq\langle A_{11}\rangle_{\rho*} \langle A_{12}\rangle_{\rho*} \langle A_{13}\rangle_{\rho*}=0.
\end{equation}
In such extreme case, the Passerby Observers essentially acts like a $t$-design \cite{4262758},  preparing a new state between each measurement, turning the measurement round into three independent measurements. We can see this in Eq.~\eqref{eq: three time prob}, where the application results in the outcome probability distribution of each measurement becoming independent of the others.

Although this analysis allows us to glimpse how noncontextuality might emerge from our PS, particularly in the classical limit \(N\rightarrow \infty\), it does not tell the whole story. We can also understand this process by moving our focus from the state to the measurements and their compatibility structure. For that, we need to analyze the situation in the Heisenberg picture. To do this, we first calculate the adjoint of channel \(\Gamma\)~\eqref{canal}, which is simply
\begin{align}
    \Gamma^*(\cdot) = \frac{1}{2}(\cdot)+\frac{1}{18}\sum_{ij}A_{ij}(\cdot) A_{ij}.
    \label{eq:DualChannel}
\end{align}

From the commutation relations of the PM observables~\eqref{squareMeasurements}, we know that
\begin{align}
  A_{ij} A_{kl} A_{ij} =
    \begin{cases}
      A_{kl} & \text{if $k=i$ or $l=j$};\\
      -A_{kl} & \text{otherwise}.
    \end{cases}   
\end{align}
Substituting this in Eq.~\eqref{eq:DualChannel}, we see that, in a system with one Passerby Observer, the Main Observer's measurements suffer a shrinking effect:
\begin{align}
    \Gamma^*(A_{kl})=\frac{5}{9}A_{kl}.
\end{align}
This Heisenberg channel provides an alternative explanation for the loss of SIC: In the Heisenberg picture, we eliminate the in-between-measurement dynamics that may destroy the correlations, but the measurements themselves change. Since SIC is sensitive to the measurements, one may not witness KS-contextuality\footnote{This results seem even more plausible when one remembers that depolarized Pauli measurements can become jointly measurable \cite{guerini2017operational}, and collective joint measurability brings forth KS-noncontextuality.}.

Generalizing to a system with $N$ Passerby Observers, we can predict the value of $\Sigma$ obtained by the Main Observer. Let us again take, as an example, context $R_1=\{A_{11},A_{12},A_{13}\}$. After many measurement rounds, as depicted in Fig.~\ref{fig: open system}, they get an average value of
\begin{align}
\begin{split}
    &\langle A_{11}\Gamma^{*N}(A_{12})\Gamma^{*N}(A_{13})\rangle_{\rho^{(0)}}=\\
    &=\left(\frac{5}{9}\right)^{2N}\langle A_{11}A_{12}A_{13}\rangle_{\rho^{(0)}}=\left(\frac{5}{9}\right)^{2N},
    \label{eq: average value}
    \end{split}
\end{align}
regardless of the states \(\rho^{(0)}\). The result is the same for all contexts of the PM square. Consequently,  given a sufficiently large number of rounds, a Main Observer evaluating Ineq.~\eqref{inequality} in a Public System with $N$ Passerby Observers would find a value of
\begin{align}
\label{fit}
    \Sigma\approx6\left(\frac{5}{9}\right)^{2N}.
\end{align}
Comparing these analytical results with the numerical simulations, as shown in Fig.~\ref{PPS-figure}, we see that the values from Eq.~\eqref{fit} perfectly match those obtained from the simulation.

Note that, in the Heisenberg picture, the Main Observer's measurements become unsharp, and KS-contextuality inequalities are no longer suitable to test the failure of a noncontextual model for the statistics (roughly, no-violations can be a consequence of a justified failure of outcome determinism)~\cite{Spekkens_Contextuality2005,Spekkens_2014_OoutcomeDeterminsmInC,Krishna_2017_PMNoiseRobust}. Nevertheless, these calculations give us an alternative explanation for the disappearance of violations: the effectively implemented measurements are not those needed to obtain the violations of the original SIC inequality.

\section{Discussion and Perspectives}

\label{Discussion}

Ref.~\cite{Baldijao_2020_StateDependentMultiple} showed that state-dependent contextuality, manifest through the odd $N$-cycle inequalities, will not survive a multiple observers setup. Due to other observers' measurements, state degradation could be responsible for the phenomenon. This hypothesis immediately suggests that state-\emph{in}dependent contextuality should be immune to such a process.

Perhaps surprisingly, we have shown that even SIC cannot resist a multiple observers' setup. We first obtained this result via a numerical simulation of numerous rounds of PM measurements in a Public System. We then discuss analytical formulations of the problem to better understand these results.

Firstly, in the Schrödinger picture, we showed that the in-between measurements act like a depolarization channel. As outlined in Appendix~\ref{Appendix}, the transformation of the state between sequential measurements is responsible for establishing the correlations among their outcomes. Introducing an additional channel between the measurements disrupts the correlations and, in the classical limit, renders the Main Observer's measurements completely independent. This result points to an important distinction: SIC is not sensitive to the state (or whatever happens to it) before the system enters the first measurement or after the last measurement of a context. SIC is, however, extremely sensitive to what happens between the measurements of a context. 

When we look at the Heisenberg picture, we can show how the in-between measurements have a shrinking effect on the Main Observer's measurements. This effect provides an alternative explanation for why violations might disappear: the measurements effectively implemented on the state are not those the Main Observer expected to implement. The Heisenberg picture also allowed us to straightforwardly predict the value of \(\Sigma\) obtained by the Main Observer as a function of the number of Passerby Observers, which perfectly matched the ones obtained from the simulation.

Finally, the analysis within the Heisenberg picture also emphasizes that contextuality is not solely a resource residing in the state; to comprehend it fully, we should also consider the measurements. This work then highlights the necessity of a KS-contextuality resource theory that emphasizes channels and multi-time measurements rather than exclusively on the state \cite{liu2019resource, Berk2021resourcetheoriesof}.

Let us comment on the relation of our work to the other results regarding contextuality in PS-like setups. Ref.~\cite{Anwer2021noiserobust} reports that \emph{preparation contextuality} can be shared between an arbitrary number of sequential observers. This result is not contradictory to the emergence of KS-non-contextuality we find here, for at least two reasons: first, preparation contextuality and SIC are different concepts -- the first related to the generalized notion of contextuality due to Spekkens~\cite{Spekkens_Contextuality2005} while the latter is based on the Kochen-Specker notion. Second, the arbitrary sharing of preparation contextuality is found by adequately adjusting the system's dimension and each observer's measurements, taking advantage of unsharpness to diminish the state degradation. This adjustment implies some shared strategy between the observers, even if this strategy is agreed upon before the measurements begin. Here, however, we consider a fixed protocol, in which one Main Observer tries to violate the Peres-Mermin inequality by measuring a system with sequential public access, with no \textit{a priori} agreement with the Passerby Observers.

The phenomenon we describe here is present,  even if subtly concealed, in Ref.~\cite{Wajs_2016_State-recycling}, which discusses the possibility of recycling the state in SIC experiments. They sequentially implement measurements of a SIC scenario chosen uniformly at random. This way of measuring also leads to interleaving incompatible measurements. However, post-processing discards all outcomes that arise from sequences that do not form a context. This data analysis procedure, as they show, leads to violations. Therefore, our work also highlights the importance of post-selection in this process; otherwise, they would not obtain violations. In our PS, however, where passerby observers are independent, post-processing is unavailable to the Main Observer, which blocks the possibility of violations.

An interesting byproduct of our results is that Eq.~\eqref{fit} can be considered in the case of a continuous parameter \(\Sigma\approx 6(5/9)^t\), which is a good way to interpret why experiments with sequential measurements of the PM square \cite{Kirchmair_2009} never reach the quantum value $\Sigma=6$: besides the error associated with the implementation of each measurement, noise introduced to the state between the measurements (which, as we have seen, is equivalent to changing the measurements themselves), scales down the violations seen in the laboratory.
They show that experimental implementations of SIC tests -- or protocols where SIC is the underlying resource-- should be careful about the interval between the measurements and the dynamics that may affect this in-between state.

There is more to explore regarding SIC in PS setups. For instance, our results suggest revisiting Mermin's proof that, if a set of measurements contain KS-sets, Vaidman's trick~\cite{Vaidman_87_XYZ} to obtain predictions for incompatible measurements does not work~\cite{Mermin_95_NoGeneralizationOfVaidmansTrick}. In a different foundational aspect, Public Systems could provide a model to analyze the disappearance of SIC due to the interaction of many quantum systems, such as shown to happen to generalized contextuality in quantum Darwinist processes~\cite{Baldijao_QDNC}. Further studies could also incorporate CP maps or varying reference frames in our analysis in Section~\ref{Results} to get a continuous model that can quantify the loss of contextuality in experimental realizations of the Peres-Mermin square, such as in Ref.~\cite{Kirchmair_2009}. From a more practical perspective, the disappearance of violations that depend on the access of third parties to the system could have exciting applications, such as the certification of private channels.

\section*{Acknowledgements}
We would like to thank Rafael Rabelo,  Matthew Pusey and Pawe\l~Horodecki for fruiful conversations.
This study was financed in part by the Coordenação de Aperfeiçoamento de Pessoal de Nível Superior (CAPES)- Finance Code 001- by CNPq under grant 310269/2019-9. This work is part of the National Institute of Science and Technology of Quantum Information (INCTIQ).
RDB acknowledges support by the Digital Horizon Europe project FoQaCiA, Foundations of quantum computational advantage, GA No. 101070558, funded by the European Union and from Instituto Serrapilheira, Chamada 2020.
\bibliographystyle{unsrtnat}
\bibliography{Bib}

\appendix

\section{Analysis of channel $\Gamma$}
\label{appendix depolarzation}
In this appendix we will show how the Main Observer perceives the channel induced by a Passerby Observer's measurements
\begin{equation}
     \Gamma( \rho)=  \frac{1}{2}\rho+\frac{1}{18}\sum_{ij}A_{ij}\rho A_{ij}
     \label{eq: canal v1}
\end{equation}
as a depolarization channel.

We can write any two-qubit density operator in the Pauli basis
\begin{equation}
\label{rho basis}
    \rho=\frac{1}{4}\left(I+\sum_{ij}t_{ij}A_{ij}+\sum_i l_iL_i +\sum_i r_iR_i \right),
\end{equation}
where
\begin{equation} 
    L_i=\sigma_i\otimes I, \quad R_i=I\otimes\sigma_i
\end{equation}
and $A_{ij}$ are the measurements from the PM square.

This decomposition allows us to see that
\begin{equation}
    \frac{\sum_{ij}A_{ij}\rho A_{ij}+4P_{L+R}(\rho)-\rho}{8}=\rho^*,
\end{equation}
where $P_{L+R}$ is the projection on the $L$ and $R$ elements, \emph{i.e.}, given $\rho$ written as in Eq.~\eqref{rho basis}
\begin{equation}
    \label{eq: channel identity}
    P_{L+R}(\rho)=\frac{1}{4}\left(\sum_i l_iL_i +\sum_i r_iR_i \right).
\end{equation}

Eq.~\eqref{eq: channel identity} allows us to rewrite channel $\Gamma$ \eqref{eq: canal v1} as
\begin{equation}
    \label{eq: real channel}
    \Gamma(\rho)=\frac{5}{9}\rho+\frac{4}{9}\rho^*-\frac{2}{9}P_{L+R}(\rho),
\end{equation} 
which differs from the depolarization channel 
\begin{equation}
\label{depol channel appendix}
    \Tilde{\Gamma}(\rho)=\frac{5}{9}\rho+\frac{4}{9}\rho^*
\end{equation} 
only on the $l_i$ and $r_i$ coefficients. We will show that, if the Main Observer is only allowed to measure PM observables, they cannot distinguish channels $\Gamma$ \eqref{eq: real channel} and $\Tilde{\Gamma}$ \eqref{depol channel appendix}.
 
When the Main Observer implements a PM measurement 
\begin{equation}
    A_{ij}=\Pi_{ij}^+-\Pi_{ij}^-
\end{equation}
on a state $\rho$, the probability of obtaining output $\pm 1$ is
\begin{align}
\begin{split}
    &p_{\rho}(\pm 1)=tr(\Pi^\pm_{ij}\rho)\\
    &=tr\left(\frac{1}{2}\left(I\pm A_{ij}\right)\rho\right)=\frac{1}{2}\pm \frac{1}{2}t_{ij},
\end{split}
\end{align}
which depends only on the $t_{ij}$ coefficient of $\rho$. But the Main Observer implements more than one measurement, so we still need to show that the state-update rule 
\begin{equation}
    \rho^{(i+1)} = \frac{\Pi_{ij}^{\pm} \rho^{(i)} \Pi_{ij}^{\pm}}{tr[\Pi_{ij}^{\pm} \rho^{(i)}  \Pi_{ij}^{\pm}]},
\end{equation}
does not propagate the disturbances on $l_i,r_i$ introduced by channel $\Gamma$ \eqref{eq: real channel} to the $t_{ij}$ components.

Considering that
\begin{equation}
    \Pi_{ij}^\pm=\frac{1}{2}\left(I\pm A_{ij}\right)
\end{equation}
we can write
\begin{align}
\begin{split}
    \Pi_{ij}^\pm(L_k)\Pi_{ij}^\pm&=\frac{1}{4}\left(L_k\pm A_{ij}L_k\pm L_k A_{ij} +A_{ij}L_kA_{ij}\right)\\
    &=\begin{cases}
        \frac{1}{2} (L_k\pm A_{ij}L_k) \text{ if } A_{ij},L_k \text{ commute;}\\
        0 \text{ if } A_{ij},L_k \text{ anticommute.}
    \end{cases}
\end{split}
\end{align}

If $L_k$ and $A_{ij}$ commute, $A_{ij}L_k$ is the third element from a context, which is also not in the PM square (see Ref.~\cite{holweck2021testing}), so
\begin{equation}
    \Pi_{ij}^\pm(L_i)\Pi_{ij}^\pm\in span(L,R),
\end{equation}
and analogously
\begin{equation}
    \Pi_{ij}^\pm(R_i)\Pi_{ij}^\pm\in span(L,R).
\end{equation}

Since the the disturbance introduced by the factor $P_{L+R}(\rho)$ does not propagate to the $t_{ij}$ coefficients of $\rho$, the Main Observer cannot differentiate the effects of channels $\Gamma$ and $\Tilde{\Gamma}$.

\section{Expected value in sequential measurements}
\label{Appendix}
The average value of a measurements $A_i$ on a state $\rho$ is:
\begin{equation}
\langle A_i\rangle_\rho \coloneqq \sum_{a_i\in \sigma(A_i)}p_\rho(a_i)a_i,
\end{equation}
with
\begin{equation}
\label{eq: prob 1 measurement}
    p_\rho(a_i)=\text{tr}[\Pi_i^{a_i}(\rho)],
\end{equation}
where $\sigma(A_i)$ is the spectrum of $A_i$ and $\Pi_i^{a_i}$ is the projection onto the $a_i$ subspace.

By fixing an state-update rule (in our case, we consider the usual textbook quantum rule, Eq.~\eqref{eq: stateUpdate}), one can define the average of measuring observable $A_i$ followed by a sequential measurement of observable $A_j$ on state $\rho$: %\sout{. The average value of sequentially measuring $A_i$ followed by $A_j$ on state $\rho$ is}
\begin{equation}
    \langle A_i\rightarrow A_j \rangle_\rho\coloneqq\sum_{\substack{a_i\in \sigma(A_i)\\ a_j\in \sigma(A_j)}}p_\rho(a_i\rightarrow a_j)a_ia_j,
\end{equation}
with
\begin{equation}
p_\rho(a_i\rightarrow a_j)\coloneqq p_\rho(a_i)p_\rho(a_j|a_i),
\end{equation}
where $p_\rho(a_j|a_i)$ is the probability of obtaining \(a_j\) from measuring $A_j$ given that \(a_i\) was obtained from measuring $A_i$.
This distribution is obtained from
\begin{equation}
\label{ai}
p_\rho(a_j|a_i)=\text{tr}[\Pi_j^{a_j}\Phi_i^{a_i}(\rho)],
\end{equation}
where $\Phi_i^{a_i}$ is a conditional channel implementing the state update rule~\eqref{eq: stateUpdate}. Notice how the dependency on $a_i$ in Eq.~\eqref{ai} is what allows for correlations between the measurements.

Eq.~\eqref{ai} reveals that, when calculating this sequential average value, the measurements are actually implemented on different states.
Therefore, we can understand this value as an average, over many runs, of the product of two outcomes: the outcome obtained from $A_i$ acting upon state $\rho$ and the outcome of $A_j$ acting upon the post-measurement state. 
It is natural to identify these values with two-point correlation functions and write:
\begin{equation}
    \langle A_i\rightarrow A_j \rangle_\rho=\langle A_i(\rho) \cdot A_j[\Phi_i^{a_i}(\rho)] \rangle.
\end{equation}
These two-point statistics are fairly common in many areas of physics, particularly in statistical mechanics (see~\cite[ch. 10]{sethna2021statistical}).

Since the sequential measurements are generally compatible and there are no disturbances between measurements, the order of implementation and state-update process is commonly omitted,
\begin{equation}
    \langle A_i\rightarrow A_j \rangle_\rho=\langle A_iA_j\rangle_\rho=\langle A_jA_i\rangle_\rho=\langle A_j\rightarrow A_i\rangle_\rho,
\end{equation}
and any of these quantities can be interpreted as an average value of the observable $A_iA_j$.
However, if we introduce a disturbance between the measurements, given by a channel $\Gamma$, the average value of the sequential measurements becomes
\begin{equation}
     \langle A_i\stackrel{\Gamma}{\rightarrow} A_j\rangle_\rho \coloneqq\sum_{\substack{a_i\in \sigma(A_i)\\ a_j\in \sigma(A_j)}}p_{\rho}(a_i\stackrel{\Gamma}{\rightarrow}a_j)a_ia_j,
\end{equation}
with
\begin{equation}
    p_{\rho}(a_i\stackrel{\Gamma}{\rightarrow}a_j)=\text{tr}[\Pi_i^{a_i}(\rho)]\text{tr}[\Pi_j^{a_j}\Gamma\Phi_i^{a_i}(\rho)].
\end{equation}
Since $\Gamma$ acts on $\Phi_i^{a_i}(\rho)$, it can modulate the correlations between the measurements.

This can be straightforwardly generalized to three sequential measurements
\begin{equation}
   \langle A_i\stackrel{\Gamma}{\rightarrow} A_j\stackrel{\Gamma}{\rightarrow} A_k\rangle_\rho \coloneqq\sum_{\substack{a_i\in \sigma(A_i)\\ a_j\in \sigma(A_j)\\ a_k\in \sigma(A_k)}}p_{\rho}(a_i\stackrel{\Gamma}{\rightarrow}a_j\stackrel{\Gamma}{\rightarrow}a_k)a_ia_ja_k,
\end{equation}
where
\begin{align}
\begin{split}
        \label{eq: three time prob}p_{\rho}(a_i\stackrel{\Gamma}{\rightarrow}a_j\stackrel{\Gamma}{\rightarrow}a_k)=&\text{tr}[\Pi_i^{a_i}(\rho)]\text{tr}[\Pi_j^{a_j}\Gamma\Phi_i^{a_i}(\rho)]\\
        &\times\text{tr}[\Pi_k^{a_k}\Gamma\Phi_j^{a_j}\Gamma\Phi_i^{a_i}(\rho)],
\end{split}
\end{align}
 and we can still identify this with a three-point correlation function
\begin{align}
\begin{split}
\label{eq: Three point correlation}
    &\langle A_i\stackrel{\Gamma}{\rightarrow} A_j\stackrel{\Gamma}{\rightarrow} A_k\rangle_\rho\nonumber\\
    &=\left\langle A_i[\rho]\cdot A_j[\Gamma\Phi_i^{a_i}(\rho)] \cdot A_k[\Gamma\Phi_j^{a_j}\Gamma\Phi_i^{a_i}(\rho)]\right\rangle.    
\end{split}
\end{align}

\end{document}